\documentclass[twocolumn,showpacs,preprintnumbers,amsmath,amssymb,superscriptaddress]{revtex4-1}
\usepackage{graphicx}% Include figure files
\usepackage{dcolumn}% Align table columns on decimal point
\usepackage{bm}% bold math
\usepackage{latexsym}
\renewcommand{\vec}[1]{\boldsymbol{#1} }

\begin{document}

\title{Internal and surface waves in vibrofluidized granular materials: Role of cohesion}
\author{Kai Huang} 
\email{kai.huang@uni-bayreuth.de}
\affiliation{Experimentalphysik V, Universit\"at Bayreuth, 95440 Bayreuth, Germany}
\date{\today}

\begin{abstract}

Wave phenomena in vibrofluidized dry and partially wet granular materials confined in a quasi-two-dimensional geometry are investigated with numerical simulations considering individual particles as hard spheres. Short ranged cohesive interactions arising from the formation of liquid bridges between adjacent particles are modeled by changing the velocity dependent coefficient of restitution. Such a change effectively suppresses the formation of surface waves, in agreement with previous experimental observations. The difference in pattern creation arises from the suppressed momentum transfer due to wetting and it can be quantitatively understood from an analysis of binary impacts.

\end{abstract}

\maketitle

\section{Introduction}
\label{intro}

Due to dissipative particle-particle interactions, energy injection is essential for exploring granular dynamics from both physical and engineering perspectives~\cite{Jaeger1996,Duran2000,Aranson2009}, as well as for handling granular materials in widespread applications ranging from geo-technique to chemical engineering~\cite{PG2017}. Depending on specific configurations, granular flow may be driven by gravity as in a silo~\cite{Janssen1895} or in planetary formation~\cite{Blum2008}, by interstitial fluid drag~\cite{Charru2013,Bagnold2005,Groh2010,Huang2012} or by a boundary~\cite{Iveson2001,Huang2010,May2013}. Consequently, the balance between the energy injection and dissipation gives rise to various nonequilibrium steady states (NESS) that pose challenges in understanding granular materials as a continuum~\cite{Goldhirsch2003,Brilliantov2004,Jenkins1983}. Note that such a balance, particularly in the boundary driven case, is rarely established instantaneously and wave propagation plays an essential role in redistributing the injected energy~\cite{Aoki1995,Bougie2002,Huang2006}. For cohesionless dry granular materials under vibrations, theoretical, numerical and experimental investigations have shown that mechanical perturbations evolve into shock waves with abrupt changes of granular temperature, pressure and density~\cite{Goldshtein1996,Bougie2002,Huang2006,Sano2011}. Furthermore, surface instability~\cite{Huang2005,Bougie2010,Almazan2013,Zippelius2017}, granular flow over obstacles~\cite{Amarouchene2006,Boudet2013}, convection~\cite{Klongboonjit2008,Fortini2015} and, last but not least, sedimentation~\cite{Almazan2015}, have all been found to be associated with wave propagation in granular materials. 

However, much less is known about how mechanical perturbations propagate in partially wet granular materials that are often encountered in nature (e.g., soil), industries (e.g., granulation process) and our daily lives (e.g., sand sculptures)\cite{Huang2014}. Here, partially wet refers to the situation with the liquid added being distributed as capillary bridges that bind individual particles together (e.g., wet sand on the beach). In this situation, typical liquid content (i.e., volume of the liquid over the total volume occupied by the partially wet granular sample) is only a few percent~\cite{Huang2014}. Recent experimental investigations have shown that the collective behavior of partially wet granular materials is dramatically different in comparison to their dry counterparts concerning rheological behavior~\cite{Schwarze2013}, rigidity~\cite{Scheel2008,Huang2009a}, phase transitions~\cite{Fingerle2008,Huang2009b,May2013,Huang2015}, clustering~\cite{Huang2012,Ramming2017}, and pattern formation~\cite{Huang2011,Butzhammer2015,Zippelius2017}. For instance, the surface waves reminiscent of the Faraday instability in a Newtonian fluid~\cite{Melo1994,Melo1995,Clement1996} are completely suppressed in vibrofluidized wet granular layers. Instead, period tripling kink-wave fronts dominate~\cite{Huang2011,Butzhammer2015}. This comparison suggests that the collective motion of granular materials is strongly influenced by `microscopic' particle-particle interactions. However, it is still unclear how the `micro-' and `macroscopic' scales are connected with each other.

From a `microscopic' perspective, recent investigations show that the energy loss associated with wet particle-particle interactions is mainly induced by capillary interactions, viscous drag force and inertia of the liquid film covering the particles~\cite{Antonyuk2009,Mueller2011,Gollwitzer2012}. The normal coefficient of restitution (COR) that characterizes the energy loss in wet particle impacts can be predicted analytically~\cite{Mueller2016}. Following this progress, it is intuitive to implement it in numerical simulations for predicting the collective behavior of partially wet granular materials. In comparison to force-based discrete element method (DEM) simulations~\cite{Cundall1979,Zhu2008}, liquid-mediated particle-particle interactions are treated as instantaneous events with the rebound velocities predicted by the COR of wet particle impacts. As the energy loss associated with particle-particle interactions is captured by the COR, an appropriate choice of the COR model is essential for an accurate prediction of the collective behavior in particulate systems. 

Using an event-driven (ED) algorithm~\cite{Rapaport1980} that predicts the collision event based on particle trajectories, pattern formation in vibrofluidized dry granular materials were reproduced successfully~\cite{Bizon1998}. Note that the conflict between instantaneous events assumed in the simulation and the finite collision time in reality may lead to unrealistic outcome, such as the inelastic collapse (i.e., diverging number of collisions within a finite time)~\cite{Luding1998,Mueller2012}. Nevertheless, such kind of challenges can be handled with an appropriate implementation of ED algorithms~\cite{Luding1998}. Moreover, it has recently been shown that collisions of soft, frictionless particles can also be modeled by introducing additional coefficients that account for the softness of particles~\cite{Mueller2013}. 

Here, I use an ED algorithm to predict the collective behavior of vibrofluidized granular materials based on the velocity dependent COR. Focusing on hard spheres confined in quasi-two-dimensions (Q2D), I show that a modification of the COR from the dry~\cite{Brilliantov2004} to the wet~\cite{Mueller2016} case effectively suppresses the formation of surface waves, and this change of collective behavior can be traced down to the inhibited momentum transfer at the individual particle level. 

\section{Methods}
\label{meth}

In the numerical set-up, $N$ spherical particles are confined in a monolayer of size $L_{\rm x} \times L_{\rm z}$ delimited by flat hard walls and gravity $\vec g= -g \vec e_{z}$ points in the negative $z$ direction. The particles have two translational degrees of freedom in the $x$ and $z$ directions and one rotational degree of freedom about the perpendicular $y$ axis. They are monodisperse with a diameter $d$ and mass $m$. The simulation box is driven sinusoidally in the $z$ direction; that is, its bottom and lid move according to $z_{\rm b} = A \sin(2\pi f t)$ and $z_{\rm l} = A \sin(2\pi f t) + L_{\rm z}$ respectively with vibration amplitude $A$ and frequency $f$ as control parameters. A related control parameter is the dimensionless peak acceleration ${\rm \Gamma}=4\pi^2f^2A/g$, which is associated with the maximum force acting on the granular layer. 

\begin{figure}
\centering
\resizebox{0.85\hsize}{!}{\includegraphics{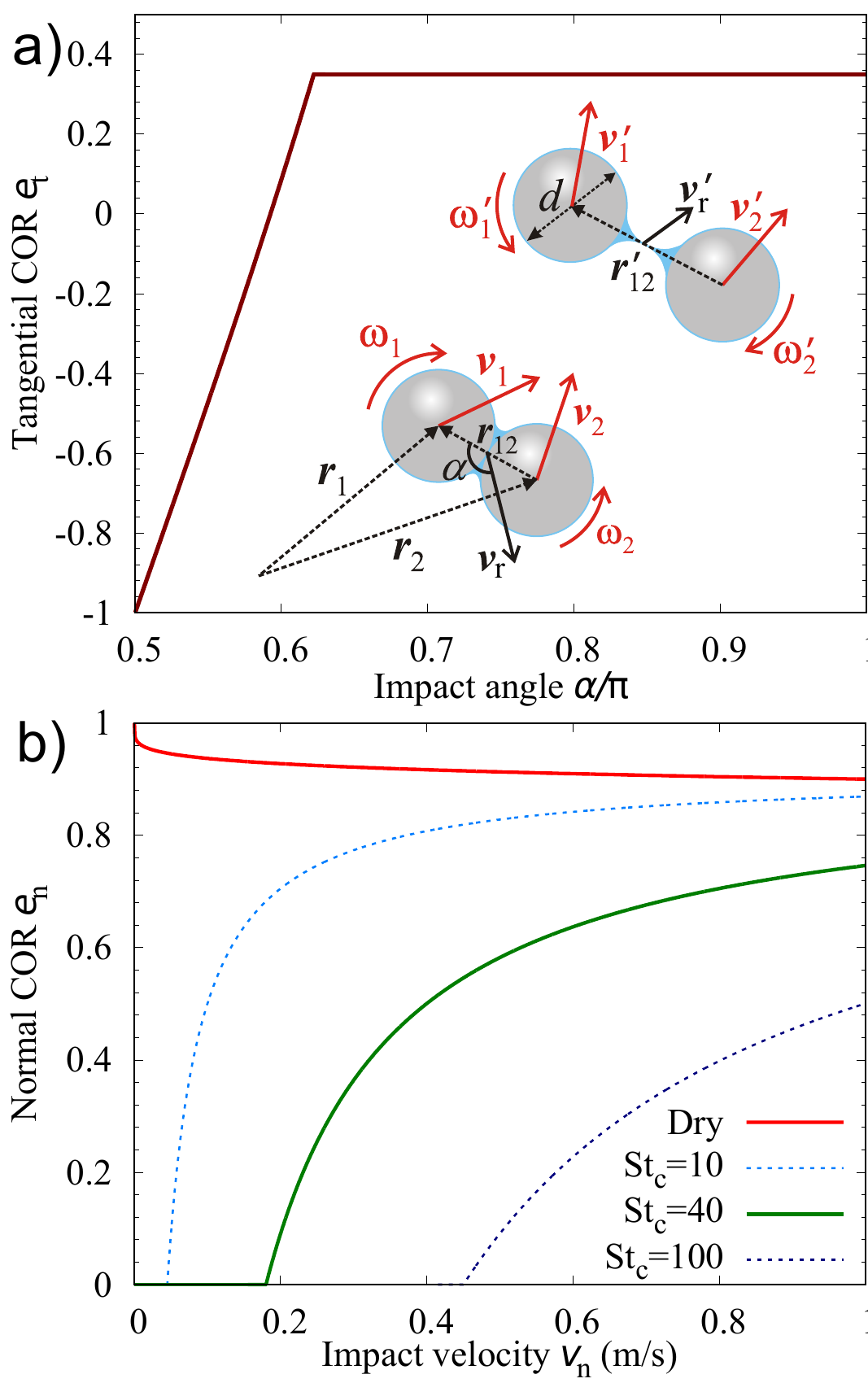}}
% \vspace{5cm}       % Give the correct figure height in cm
\caption{(Color online)(a) Tangential COR as a function of impact angle $\alpha$ for $\mu = 0.5$ and $e_{\rm t0} = 0.35$. Inset of (a) is a sketch of a binary collision with the definitions of collisional parameters. (b) Normal coefficient of restitution (COR) as a function of normal impact velocity for dry and wet particle impacts following Eq.\,\ref{eq:drycor} and \ref{eq:wetcor}, respectively. For the wet case, the normal COR for various ${\rm St}_{\rm c}$ at a fixed $e_{\rm inf}=0.91$ is presented to show the influence of this wetting parameter on the impact velocity dependent COR.}
\label{fig:cor}       
\end{figure}

After initialization, the particles are randomly located inside the simulation box with a small random velocity $< 10^{-2}\sqrt{gd}$. Subsequently, the particles fly freely under gravity until a collision event occurs. The positions, translational and angular velocities of all particles are recorded every $1/(Mf)$ second with $M=100$ the number of recorded phases per vibration cycle. As sketched in the inset of Fig.~\ref{fig:cor}(a), the relative velocity at the contact point of two colliding particles with positions $\vec r_{1,2}$, velocities $\vec v_{1,2}$ and angular velocities $\vec \omega_{1,2}$ is calculated with   

\begin{equation}
\vec v_{\rm r}=\vec v_{1} - \vec v_{2} - \frac{d}{2}(\vec \omega_{1} + \vec \omega_{2}) \times \hat{\vec n} \,
\end{equation}

\noindent where $\hat{\vec n}=\vec r_{12}/|\vec r_{12}|$ with $\vec r_{12}=\vec r_1-\vec  r_2$ corresponds to the unit vector in the normal direction. 

Based on momentum conservation, post-collisional velocities are~\cite{Walton1986,Luding1995}

\begin{eqnarray}
\vec v_{1}' = \vec v_{1} + {\rm \Delta} \vec p/m \\ \nonumber 
\vec v_{2}' = \vec v_{2} - {\rm \Delta} \vec p/m \\ \nonumber 
\vec \omega_{1,2}' = \vec \omega_{1,2} - \frac{d}{2I} \hat{\vec n} \times {\rm \Delta} \vec p \,
\end{eqnarray}

\noindent with the moment of inertia of a sphere $I=m d^2/10$. Primed variables correspond to post-collisional quantities. The momentum exchange upon impact is

\begin{equation}
{\rm \Delta} \vec p = -\frac{1}{2}(1+e_{\rm n})m \vec v_{\rm n} - \frac{\beta}{2}(1+e_{\rm t})m \vec v_{\rm t} \,
\end{equation}

\noindent where $\vec v_{\rm n} = \vec v_{\rm r} \hat{\vec n}$ and $\vec v_{\rm t} = \vec v_{\rm r} -\vec v_{\rm n}$ correspond to the normal and tangential component of $\vec v_{\rm r}$, $e_{\rm n}=-\vec v_{\rm r}' \hat{\vec n}/(\vec v_{\rm r} \hat{\vec n})$ and $e_{\rm t}=\vec v_{\rm r}' \hat{\vec t}/(\vec v_{\rm r} \hat{\vec t})$ denote the COR in the normal and tangential directions, respectively. The unit vector in tangential direction is $\hat{\vec t}=\vec v_{\rm t}/|\vec v_{\rm t}|$. The factor $\beta = 2/7$ is due to the fact that a change of the translational velocity is coupled to that of the angular velocity and the coupling factor relies on the moment of inertia of the spherical particles considered here~\cite{Luding1995}.  

As shown in Fig.~\ref{fig:cor}(b), the dependence of $e_{\rm n}$ on the magnitude of normal impact velocity $v_{\rm n}$ differs qualitatively between dry and wet particles. For dry COR $e_{\rm dry}(v_{\rm n})$, a smaller impact velocity $v_{\rm n}$ leads to a higher normal COR because of the higher tendency for the particles to interact elastically. For the wet case, an onset velocity $v_{\rm c}$, below which no rebound occurs, arises from capillary and viscous drag forces imposed by the wetting liquid~\cite{Gollwitzer2012}. Due to the dramatic difference of $e_{\rm n}(v_{\rm n})$, it is intuitive to use the velocity dependent COR for predicting the different collective dynamics between dry and wet granular materials.

More specifically, dry COR is estimated with~\cite{Brilliantov2004}

\begin{equation}
\label{eq:drycor}
e_{\rm dry} = 1 - \kappa v_{\rm n}^{1/5}
\end{equation}

\noindent with the material parameter $\kappa = 0.1$. The wet COR, on the other hand, is estimated with~\cite{Mueller2016} 

\begin{equation}
\label{eq:wetcor}
e_{\rm wet} = e_{\rm inf} [ 1 - {\rm St}_{\rm c}/(q v_{\rm n}) ]
\end{equation} 

\noindent for $v_{\rm n} \ge v_{\rm c}$ and $0$ elsewhere, where the two control parameters $e_{\rm inf}$ and St$_{\rm c}$ correspond to the COR at infinitely large $v_{\rm n}$ and the onset velocity $v_{\rm c} = {\rm St}_{\rm c}/q$, respectively. The constant factor $q = \rho_{\rm p} d/(9 \eta)$ with particle density $\rho_{\rm p}$ and dynamic viscosity $\eta$ is related to the definition of the Stokes number St$ = q v_{\rm n}$, which measures particle inertial over viscous drag force. The same collision rule is implemented for particle-wall collisions. In comparison to previous numerical simulations~\cite{Fingerle2008,Ulrich2009,Roeller2011,Schwarze2013,Roy2017} of wet granular materials that include predominately capillary interactions, the influence of inertial and viscous drag forces from the liquid film on the COR is also included here. 

Since the momentum transfer in the tangential direction is coupled to that in the normal direction by the laws of friction, the tangential COR is estimated with $e_{\rm t} = -1 - \mu (1 + e_{\rm n}) \cot \alpha / \beta $ in case of $e_{\rm t} < e_{\rm t0}$ and $e_{\rm t0}$ elsewhere [see Fig.~\ref{fig:cor}(a)], where $\mu$ is the frictional coefficient and $e_{\rm t0}$ is the limiting tangential COR introduced to account for the onset of sliding~\cite{Luding1995}. For particle-wall collisions, the tangential COR is set to $1$.  

In addition, the TC model is implemented to avoid inelastic collapse of the granular layer while colliding with the container~\cite{Luding1998}. Here, TC stands for a collisional time scale. The model considers that only the first collision of each particle within this time scale is inelastic and all subsequent ones are elastic (i.e., $e_{\rm n} = e_{\rm t} = 1$). Here, $T_{\rm c}$ is chosen to be the contact duration of two colliding elastic spheres $T_{\rm c} = 2.94( m/k)^{2/5} v_{0}^{-1/5}$ with a normal colliding velocity $v_{0}$ and material parameter $k$. The former variable is chosen to be the maximum velocity of the driving plate $v_{0}={\rm \Gamma} g/(2\pi f)$, and the latter one is calculated with $k=\frac{4E\sqrt{d}}{15(1-\sigma^2)}$ with $E$ and $\sigma$ Young's modulus and Poisson's ratio of the particle~\cite{Duran2000}. The simulation parameters are shown in Table~\ref{tab:properties}. They correspond to the case of water wetting glass spheres. Based on a previous investigation~\cite{Gollwitzer2012}, ${\rm St}_{\rm c}$ and $e_{\rm inf}$ are chosen to be $40$ and $0.91$ throughout the article unless otherwise stated. Note that for other particle-liquid combinations, the two parameters can be estimated analytically with a wet impact model~\cite{Mueller2016}. For each parameter set, a simulation time of at least $15$\,s is chosen. 

\begin{table}
\caption{\label{tab:properties}%
Numerical values of the simulation parameters.}
\begin{ruledtabular}
\renewcommand{\arraystretch}{1.3}%
\begin{tabular}{cc|c}
\textrm{Parameters}&&
\textrm{Value}\\
\colrule
 Particle density & $\rho_{\rm p}$ & 2580 kg\,m$^{-3}$ \\
 Particle diameter & $d$ & 0.002 m \\
 Young's modulus & $E$  &  63 GPa \\
 Poisson's ratio & $\sigma$  &  0.22  \\
 Container width & $L_{\rm x}$ & 30$d$ \\
 Container height & $L_{\rm z}$ & 30$d$ \\
 Dynamic viscosity of liquid & $\eta$  &  1.0 mPa\,s \\
\end{tabular}
\end{ruledtabular}
\end{table}

\section{Stability diagram}
\label{sec:pd}

\begin{figure*}
\centering
\resizebox{0.85\hsize}{!}{\includegraphics{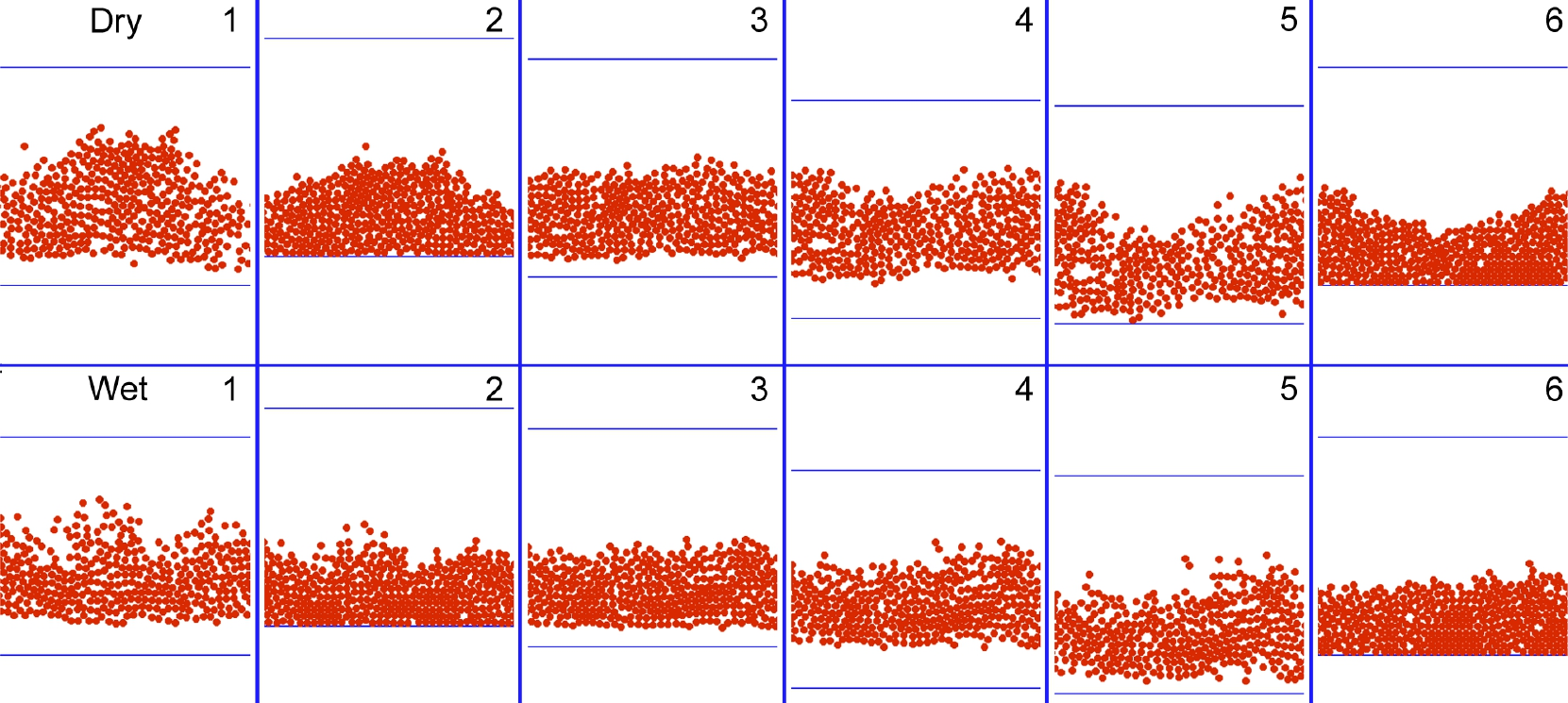}}
\caption{(Color online) Snapshots showing the collective behavior of dry (upper row) and wet (lower row) particles in one vibration cycle. Blue lines correspond to the upper lid and lower bottom of the vibrating plates. The parameters are $N=300$, $f=10$\,Hz and ${\rm \Gamma}=4$.}
\label{fig:snapshot}       
\end{figure*}

Figure~\ref{fig:snapshot} compares the collective motion of dry (upper row) and wet (lower row) particles during one vibration cycle. There exist a free-flying and a compressing regime that occurs directly after the impact of the granular layer with the container bottom for both cases. For the dry case, there is a clear tendency of forming surface waves, in agreement with a previous experiment conducted in Q2D~\cite{Clement1996}. For wet particles under the same driving conditions, surface waves are suppressed, as expected from previous investigations on wet granular materials~\cite{Huang2009b,Butzhammer2015}. An increase of $L_{\rm x}$ while keeping the same granular layer thickness and a change to periodic boundary condition in the $x$ direction yield qualitatively the same behavior. 

\begin{figure}
\centering
\resizebox{0.85\hsize}{!}{\includegraphics{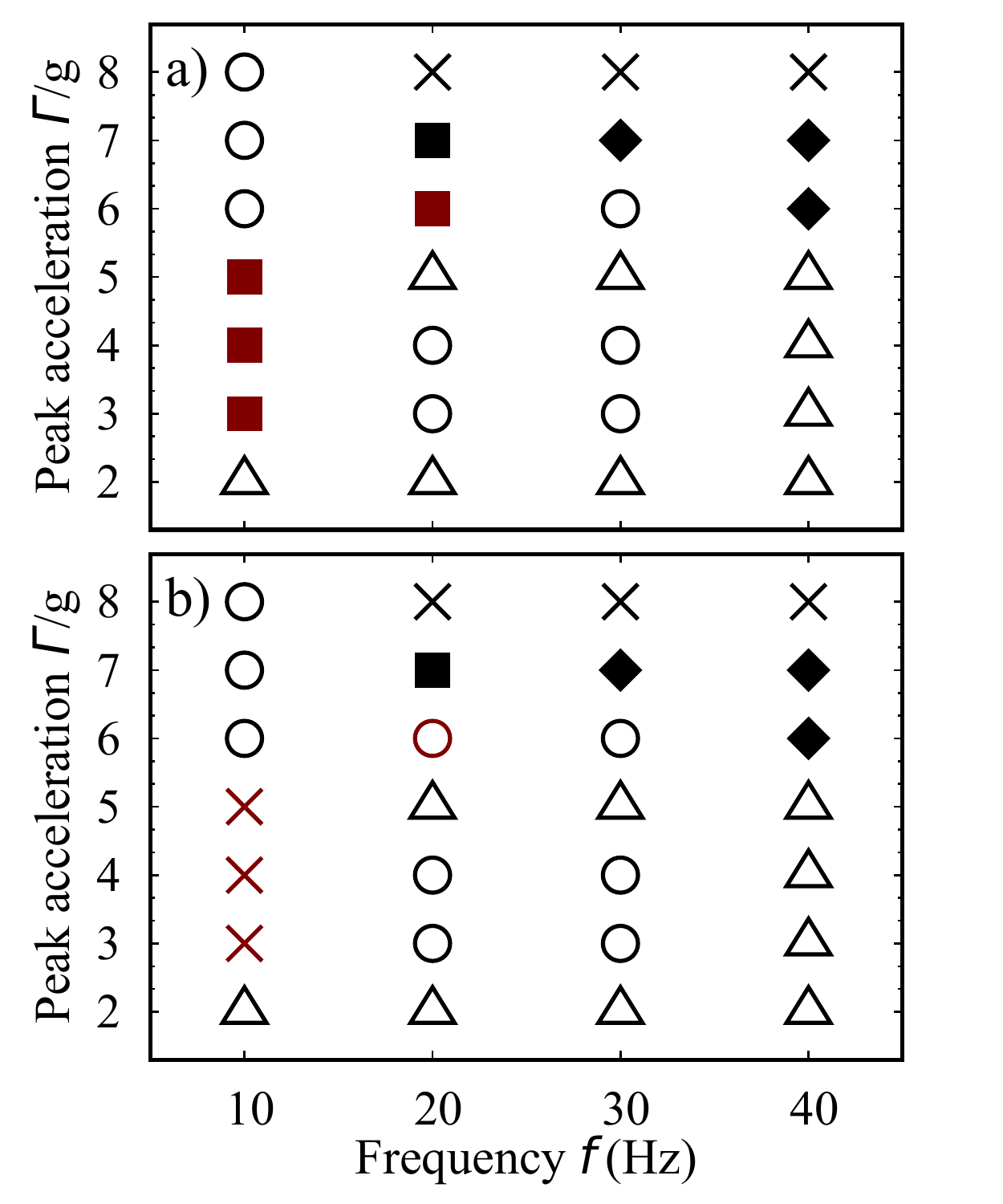}}
\caption{(Color online) Stability diagram for (a) dry and (b) wet granular layers under vertical vibrations. The $\bigtriangleup$, $\bigcirc$ and $\times$ points correspond to crystal, amorphous, and liquidlike states, respectively. The $\blacksquare$ and $\blacklozenge$ points denote surface and kink wave instabilities. Red symbols highlight the difference between the collective behavior of dry and wet particles. Other parameters are the same as in Fig.~\ref{fig:snapshot}.}
\label{fig:pd}       
\end{figure}

Figure~\ref{fig:pd} shows the stability diagram for dry (a) and wet (b) particles as a function of $f$ and ${\rm \Gamma}$. It is obtained by an inspection of the collective motion of particles during the free-flying period, after the system evolves into a steady state in which the center of mass (CM) of the granular layer fluctuates in a periodic manner. Crystalline and amorphous states are distinguished from whether the particles are caged in a crystalline or a random configuration. In a liquid-like state, the particles can move freely with respect to their neighbors. At the lowest frequency $f=10$\,Hz, an increase of ${\rm \Gamma}$ from $2$ leads to a fluidization of the granular layer for both dry and wet cases until a solidification at ${\rm \Gamma} \ge 6$, where compression due to collisions with the container lid starts. For $\Gamma \in {[3,5]}$, surface waves emerge for a fluidized dry granular layer but not for the wet case (see Fig.~\ref{fig:snapshot}). As $f$ increases, the difference between dry and wet particles becomes less prominent, because the vibration amplitude, which decays with $f^{-2}$, becomes comparable to $d$ and the free-flying time for surface waves to develop also shortens. At ${\rm \Gamma} \approx 5$, both dry and wet granular layers crystallize again because of the diminished energy injection when the impact velocity of the granular layer matches that of the vibrating plate. This corresponds to the onset of a period doubling bifurcation~\cite{Melo1995}. Note that for a wet granular layer, the bifurcation threshold is expected to vary with the cohesive force between the granular layer and the vibrating plate~\cite{Butzhammer2015}. Such a feature does not exist here as instantaneous collisions are assumed in the ED algorithm. Nevertheless, the emerging period doubling kink waves for ${\rm \Gamma} \le 6$ are in agreement with previous experiments~\cite{Melo1995,Zhang2005,Butzhammer2015}. In short, this comparison indicates that the change of velocity dependent COR captures the collective behavior of granular systems, particularly the difference in pattern creation between dry and wet granular materials.

\section{Wave propagation}
\label{sec:wave}

\begin{figure}
\centering
\resizebox{0.99\hsize}{!}{\includegraphics{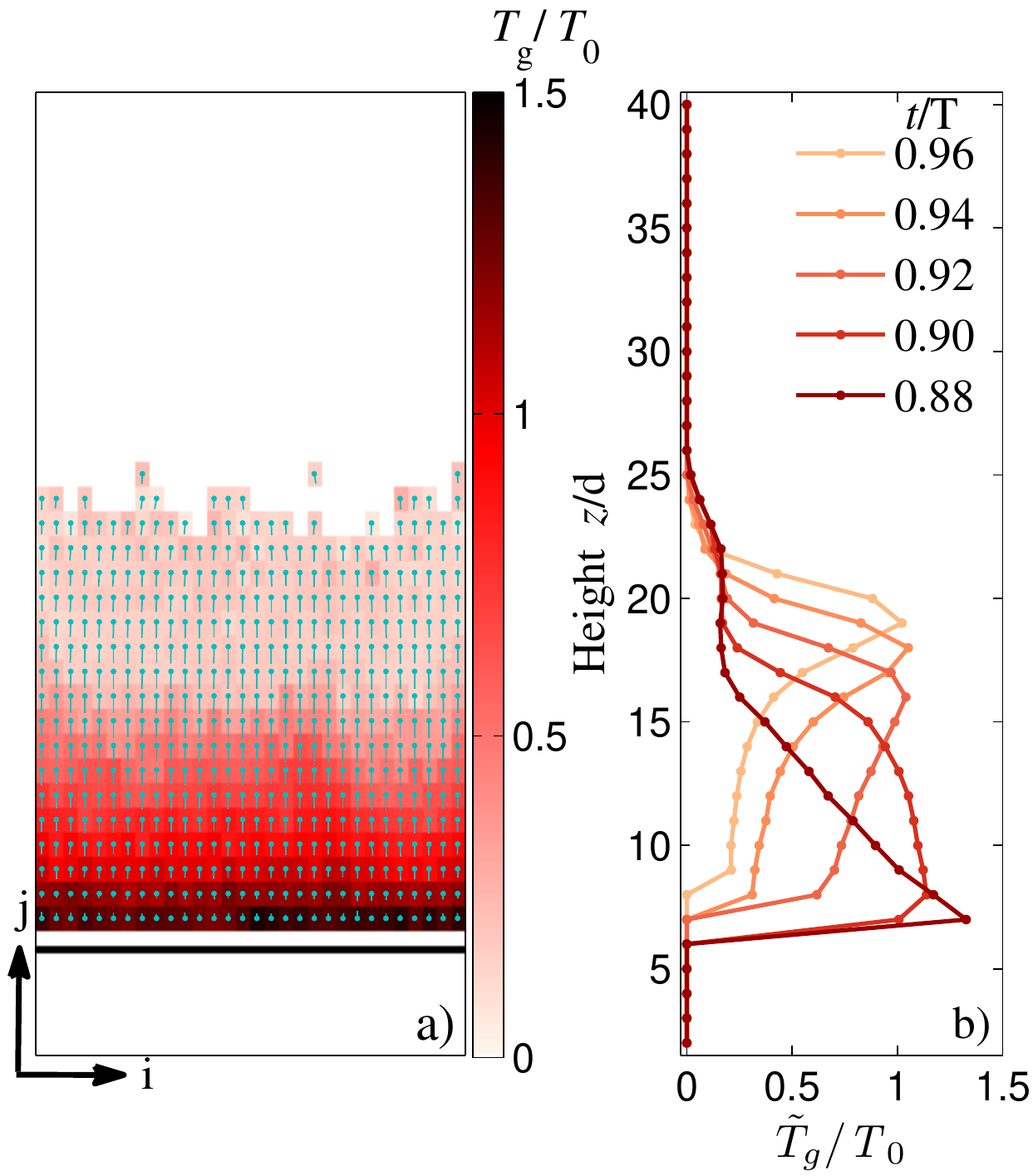}}
%\vspace{5cm}       % Give the correct figure height in cm
\caption{(Color online) Left: Velocity (denoted by arrows) and granular temperature (denoted by colors) $T_{\rm g}$ field at $t/T=0.88$ for wet particles, where $T=1/f$ is the vibration period. Solid dots correspond to the centers of the cells fixed in the lab system. The horizontal line illustrates the bottom of the vibrating container. Right: Averaged granular temperature ${\tilde T}_{\rm g}$ as a function of height in consequent time steps. Other parameters are the same as in Fig.~\ref{fig:snapshot}.} 
\label{fig:Twave}       
\end{figure}

Figure~\ref{fig:Twave} illustrates how wave fronts develop and propagate through a wet granular layer under vertical vibrations. As the free-flying period ends at $t/T=0.88$, the granular layer starts to collide with and collect energy from the vibrating plate. Consequently, a disturbed region with the mean velocity pointing upwards in the $z$ direction emerges. The counter-flow between the disturbed and undisturbed regions leads to a granular temperature peak, in which frequent particle-particle collisions occur. To obtain the spatially resolved velocity and temperature field, the space being possibly occupied by the particles is divided into $N_{\rm x} \times N_{\rm z}$ cells in the lab system, where $N_{\rm x} = L_{\rm x}/d$ and $N_{\rm z} = \lceil(L_{\rm z} +  2A)/d \rceil$ with ceiling function $\lceil \rceil$. To get better statistics, data collected at the same phase of multiple ($\ge 300$) vibrations cycles in the steady state are averaged to obtain the time-resolved velocity and granular temperature fields. In cell $(i,j)$, the granular temperature is calculated with ${T_{\rm g}(i,j) = m\sum_{k=1}^{n(i,j)}(\vec v_{k} - \tilde{\vec v}_{k})^2/[2n(i,j)]}$, where $\vec v_{k}$ corresponds to the velocity of particle $k$, $i \in {[1,N_{\rm x}]}$ and $j \in {[1,N_{\rm z}]}$ are indices of the cell, and $\tilde{\vec v}_{k} = \langle \vec v_{k} \rangle$ is obtained through an average over the velocities of all $n(i,j)$ particles inside. Only cells with sufficient number of particles $n(i,j) \ge 3$ are analyzed. The kinetic energy scale $T_{0} = m v_{0}^2/2$ is based on the maximum velocity of the vibrating plate $v_{0}$. 

As the velocity field in Fig.~\ref{fig:Twave}(a) shows, particles in the undisturbed region move collectively downwards with much less horizontal components in comparison to particles in the disturbed region just collided with the container bottom. As shown in Fig.~\ref{fig:Twave}(b), the collision results in a temperature gradient in the vertical direction being developed at $t/T=0.88$. Here the averaged granular temperature is obtained with ${\tilde T}_{\rm g}= \langle T_{\rm g}(i,j) \rangle_i$ with $\langle ... \rangle_i$ an average over different column $i$. As time evolves to $t/T=0.90$, the peak broadens while propagating upwards, indicating that intensive momentum exchange takes place throughout the granular layer. As time evolves further, the mobility of the particles in the disturbed region becomes more coherent. Subsequently, the temperature peak sharpens again until it decays while reaching the top of the granular layer. Note that it takes less than $1/10$ of a vibration period for the wave to propagate through the granular layer and redistribute the injected energy. 

\begin{figure}
\centering
\resizebox{0.95\hsize}{!}{\includegraphics{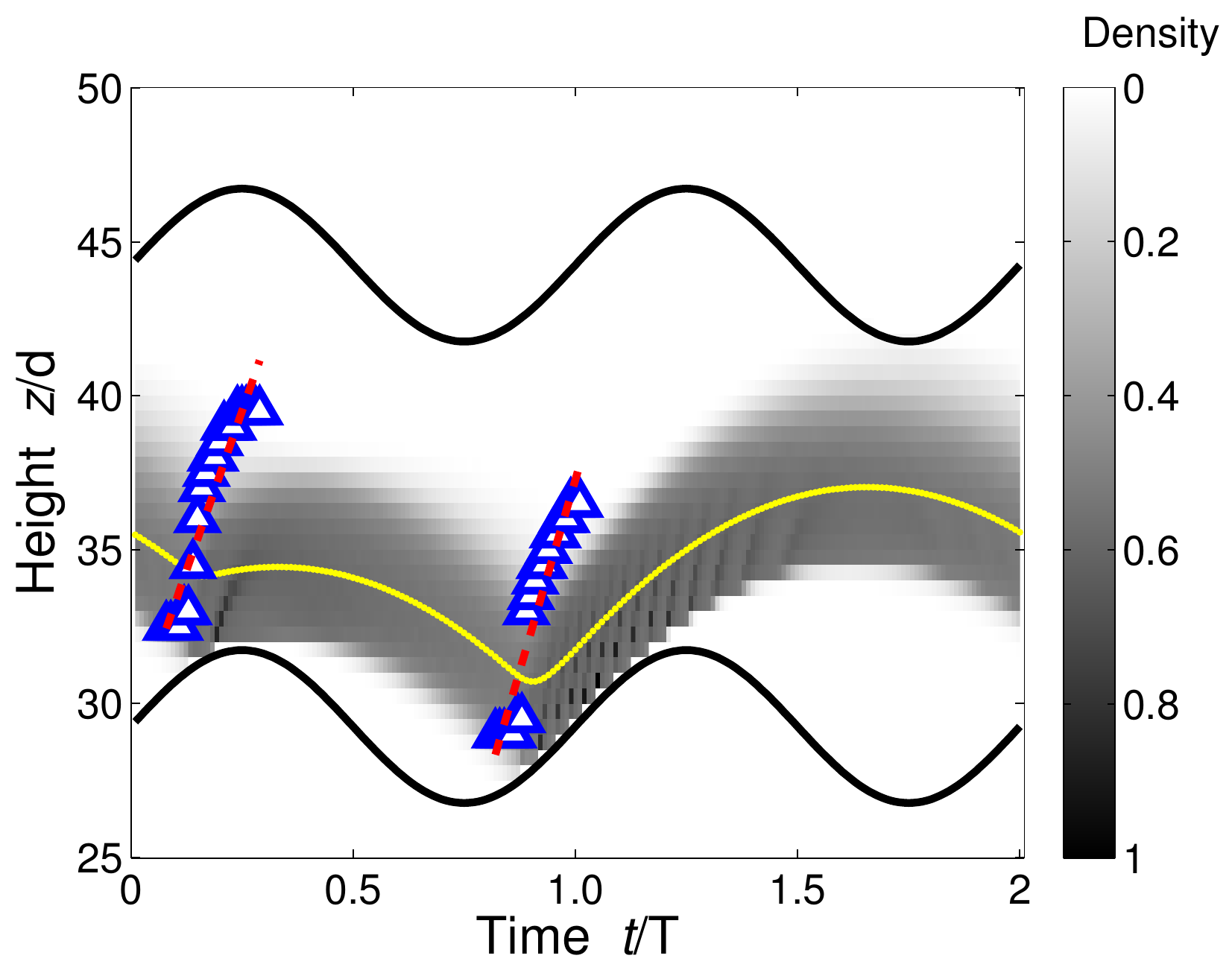}}
\caption{(Color online) Evolution of the density profile (shaded in gray) and the height of CM (yellow line) in a vibrofluidized wet granular layer in two consecutive vibration cycles. The $\bigtriangleup$ points, which represent the peak of $T_{\rm g}$ profile, show the wave propagation directly after the granular layer collides with the container bottom. Dashed lines are linear fits to the height of the wave fronts. The sinusoidal lines represent the bottom and lid of the vibrating container. Other parameters are the same as in Fig.~\ref{fig:snapshot}} 
%Upper triangle symbols
\label{fig:front}       
\end{figure}

According to previous investigations~\cite{Melo1995,Butzhammer2015}, the periodicity of both the dry and the wet granular layer under vertical vibrations can be estimated approximately with a single particle colliding completely inelastically with the container. For the driving condition used here ($f=10$\,Hz, ${\rm \Gamma}=4$), the granular layer undergoes a period doubling bifurcation and consequently the CM trajectory shown in Fig.~\ref{fig:front} has a period of $2T$. In the first vibration cycle, the granular layer detaches immediately after colliding with the container bottom (undergoing a `weak' impact), because the acceleration of the vibrating plate at collision $a < -g$. The granular layer dilutes during the free-flying period until the next collision starts. After the second collision, the granular layer stays together with the container bottom for some time before the next free-flying period starts. Wave propagation is initiated as the granular layer collides with the container bottom and the injection of kinetic energy starts. As indicated by the fitted lines, the wave propagates quickly through the granular layer before the free-flying period starts. For the `weak' impact taking place at $t/T \approx 0.1$, the wave propagates slightly slower ($1.61 \pm 0.25$\,m/s) than the one ($1.95 \pm 0.25$\,m/s) emerging after the second impact. The location of the wave front is identified as the location of a developed temperature peak (i.e., $T_{\rm g}/T_0 > 0.5$). Because of the wide range of $e_{\rm n}$ as $v_{\rm n}$ varies, it is unclear whether the sound speed predicted with existing kinetic theory~\cite{Savage1988} assuming constant COR still applies or not. Therefore, no concrete statement on the development of shock wave fronts can be made here.

\begin{figure}
\centering
\resizebox{0.95\hsize}{!}{\includegraphics{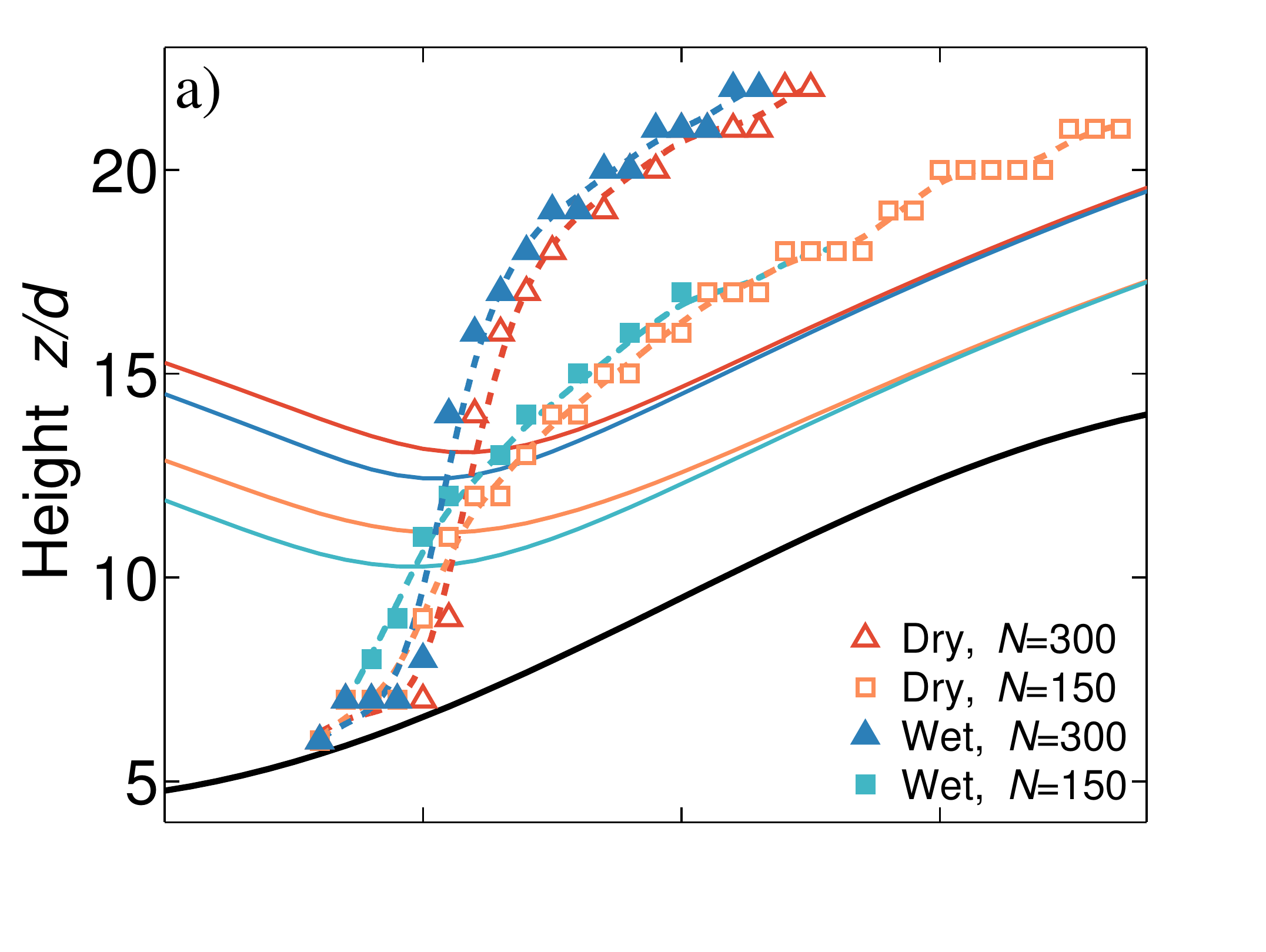}}\\
\vskip -2.5em
\resizebox{0.95\hsize}{!}{\includegraphics{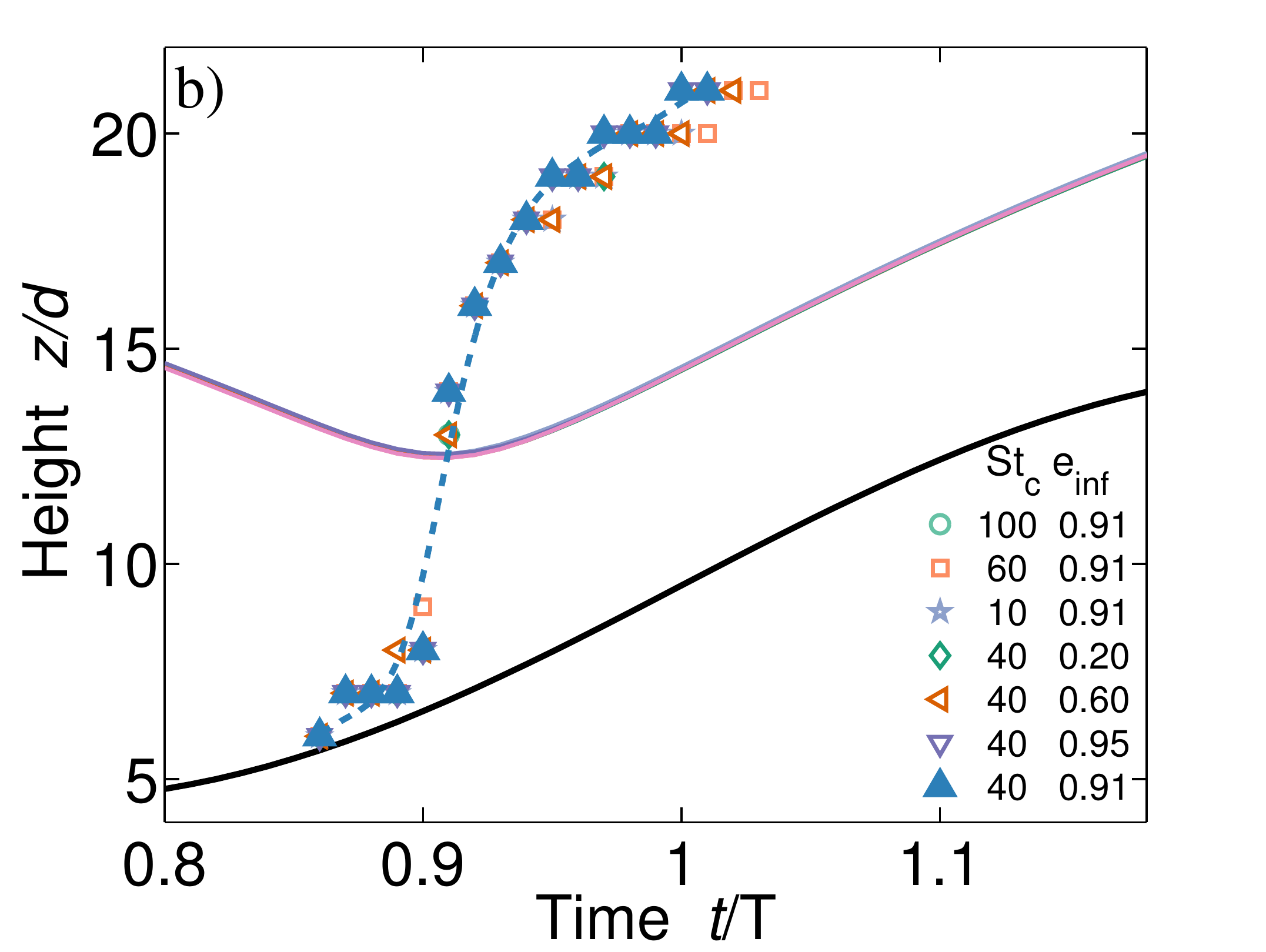}}
\caption{(Color online) (a) Comparison of wave propagation in dry and wet granular materials with different $N$. Dashed lines, which correspond to spline fits to the data, are guides to the eyes. (b) Wave propagation in wet granular layers with fixed $N=300$ and various parameters St$_{\rm c}$ and $e_{\rm inf}$. In both figures, the thin solid line of the same color as the data points represents the corresponding CM trajectory, while the thick black solid line denotes the height of the container bottom.} 
\label{fig:speed}       
\end{figure}

Figure~\ref{fig:speed}(a) shows a comparison of wave front propagation between dry and wet granular layers with different $N$. Qualitatively, the wave front propagates faster as $N$, or the number of layers increases, in agreement with previous investigations~\cite{Bougie2002,Huang2006}. This is due to the higher particle density in the disturbed region. On the contrary, the influence of wetting on the wave front propagation is relatively weak. It leads to a slightly earlier start of wave propagation, presumably due to the earlier collision of the granular layer with the container bottom. Taking the lowest point of the corresponding CM trajectory [see Fig.\,\ref{fig:speed}(a)] as a measure of the time scale of collision, we have the collision time $t/T = 0.92$ and $0.90$ for dry and wet particles with $N=300$, respectively. The delay time between dry and wet granular layer $\approx0.02T$ stays the same if we choose the onset of wave front as the order parameter. For both dry and wet granular layers, no granular Leidenfrost effect~\cite{Eshuis2007} is observed, presumably due to the limited the range of $\Gamma$ explored here.

As shown in Fig.\,\ref{fig:speed}(b), the development and propagation of wave fronts as well as the CM trajectories for various parameters $e_{\rm inf}$ and St$_{\rm c}$ overlap with each other pretty well, indicating that a quantitative modification of wet COR does not influence the propagation of the wave fronts. In comparison to Fig.\,\ref{fig:speed}(a), it is reasonable to conclude that internal wave propagation is only weakly influenced by wetting, and this influence arises from the distinct difference of $e_{\rm n}$ as $v_{\rm n}<v_{\rm c}$, not the quantitative change induced by the parameters $e_{\rm inf}$ and St$_{\rm c}$.

\section{Momentum transfer}
\label{sec:vel}

\begin{figure}
\centering
\resizebox{0.9\hsize}{!}{\includegraphics{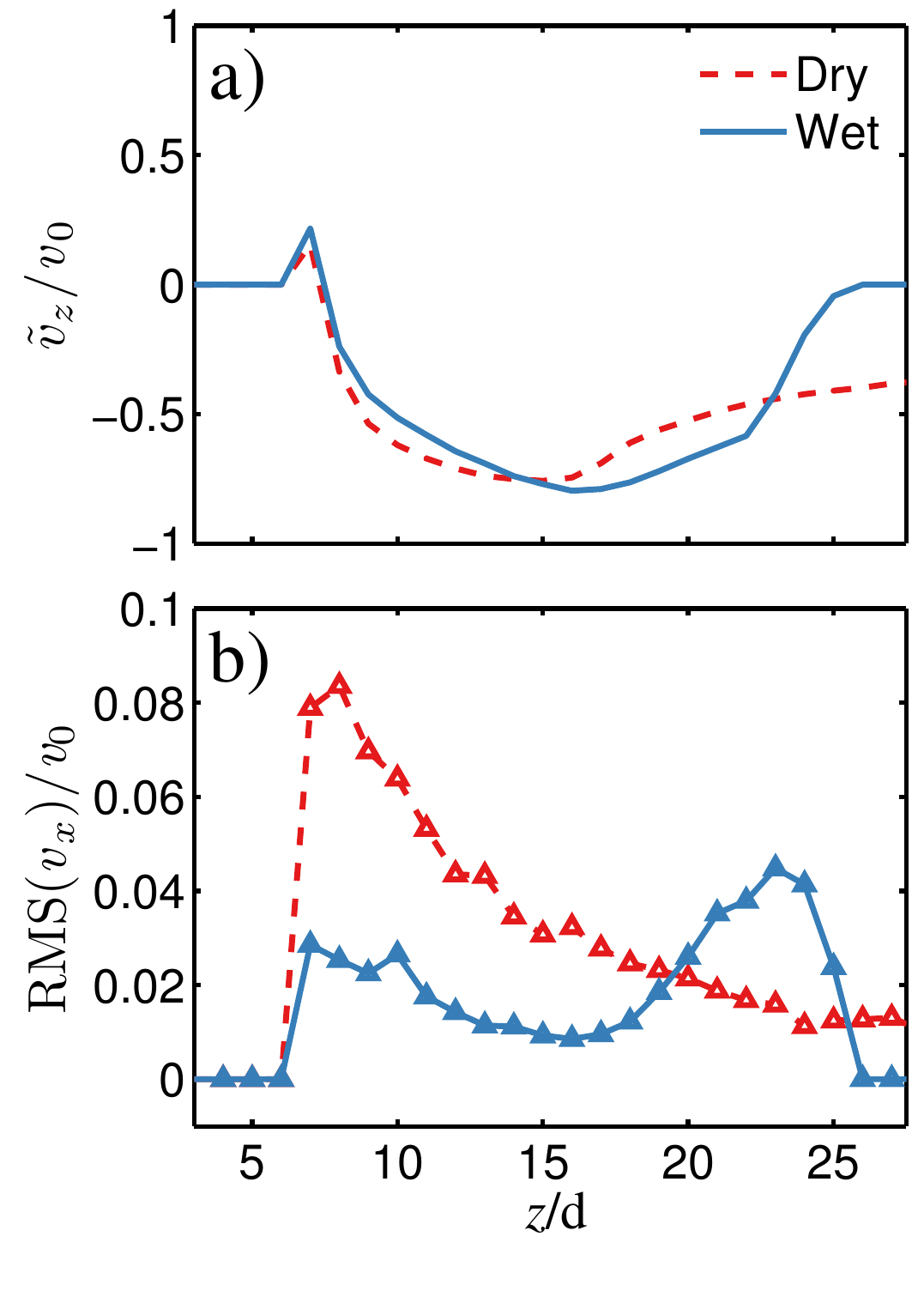}}
\caption{(Color online) Mean vertical velocity $\tilde{v}_z$ (a) and horizontal velocity fluctuations ${\rm RMS}(v_{\rm x})$ (b), both rescaled by the peak vibration velocity $v_0$, as a function of height for both dry and wet particles at $t/T = 0.88$. Other parameters are the same as in Fig.~\ref{fig:snapshot}.} 
\label{fig:vmod}       
\end{figure}

Finally, we come to the question of how the change of COR hinders the formation of standing waves. The above analysis shows that wave propagation plays a minor role as it takes place only shortly after collisions with the container bottom and depends weakly on the change of COR. For the development of standing waves, one precondition is the momentum transfer from the direction of driving to that perpendicular to driving due to frequent collisions of particles between the disturbed and undisturbed regions. From the velocity field shown in Fig.\,\ref{fig:Twave}(a), the mean vertical (i.e., driving direction) flow field can be obtained with $\tilde{v}_z = \langle  \tilde{\vec v}_k \hat{\vec z} \rangle_i$, where $\hat{\vec z}$ is the unit vector in the $z$ direction. As shown in Fig.\,\ref{fig:vmod}(a), the mean vertical velocity $\tilde{v}_z$ changes sign in the counter-flow region described above. At this moment, the vertical velocity profiles for dry and wet particles are comparable with each other. In order to quantify the momentum being transferred to the horizontal direction for surface waves, I calculate the horizontal velocity fluctuations RMS$(v_{x}) = \sum_{i=1}^{n_z}(v_{x}-\langle v_{x} \rangle_i)^2/n_z$, where $v_x = \tilde{\vec v}_k \hat{\vec x}$ is the horizontal component of the velocity field with $\hat{\vec x}$ the unit vector pointing in the $x$ direction. The order parameter RMS$(v_{x})$, which measures the modulation of the velocity field in the direction perpendicular to the driving, is expected to grow as the tendency for the creation of surface waves gets stronger. A comparison of RMS$(v_{x})$ between dry and wet granular layers in Fig.\,\ref{fig:vmod}(b) clearly indicates the suppression of momentum transfer in the counter-flow region due to wetting. As time evolves, $\tilde{v}_z$ increases as the bottom plate pushes the whole granular layer upwards, while RMS$(v_{x})$ decays correspondingly. This behavior suggests an enhanced collective motion due to driving. This comparison between dry and wet granular layer stays qualitatively the same along with the propagation of the wave front. 

\begin{figure}
\centering
\resizebox{0.85\hsize}{!}{\includegraphics{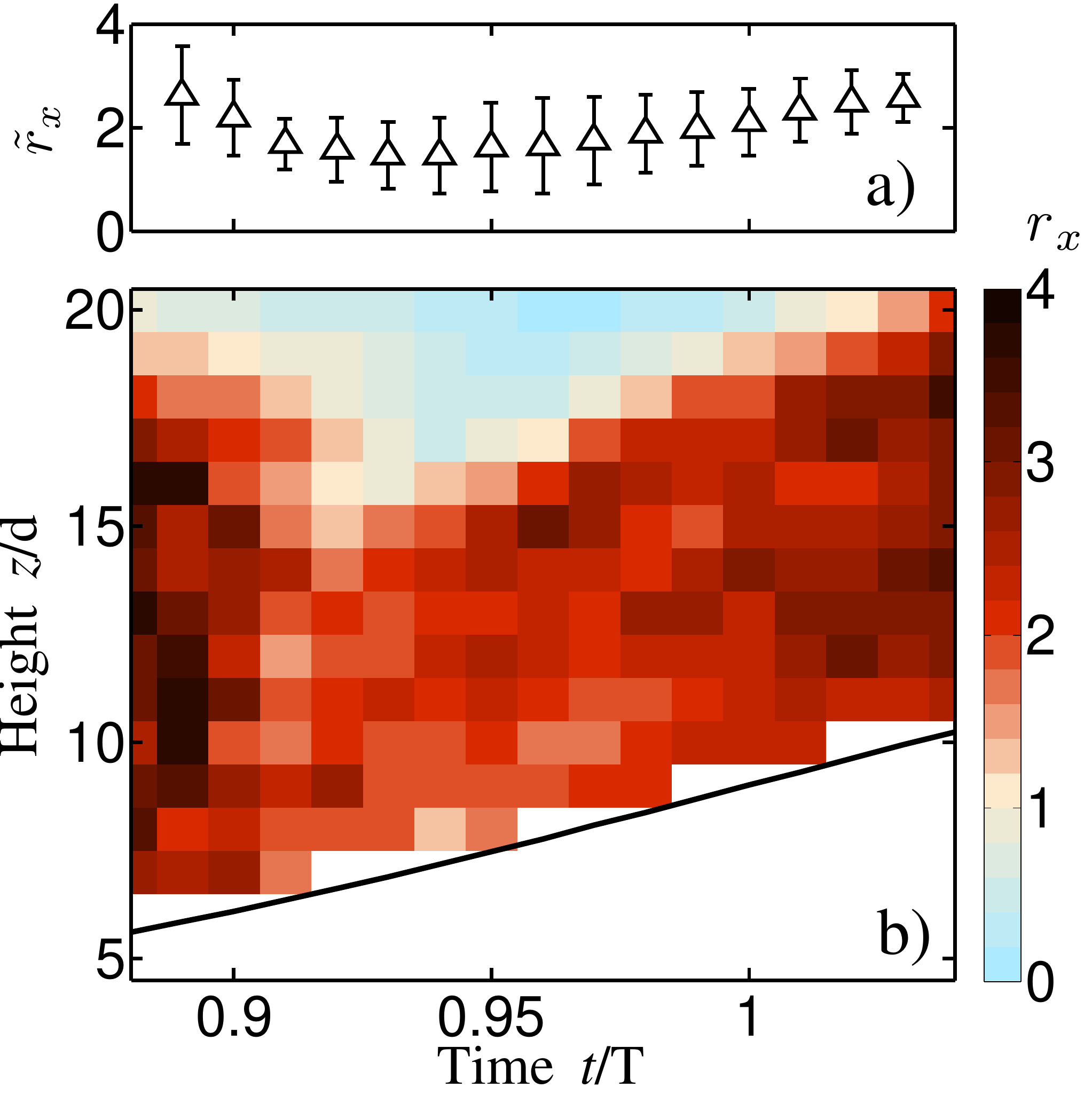}}
\caption{(Color online) (a) Mean deficiency factor $\tilde{r}_{x}$ as a function of time. (b) Time evolution of the deficiency factor $r_{x}$ as a function of height directly after the impact of the granular layer on the container bottom.} 
\label{fig:rx}       
\end{figure}

In order to quantify the difference between dry and wet granular layers, a deficiency factor of momentum transfer is defined as $r_{x}=\frac{{\rm RMS}(v_{x})_{\rm dry}}{{\rm RMS}(v_{x})_{\rm wet}}$, which compares the horizontal velocity fluctuations for dry and that for wet particles. As shown in Fig.\,\ref{fig:rx}(a), the mean deficiency factor $\tilde{r}_{x} = \langle r_{x} \rangle_{j}$ with $\langle ... \rangle_{j}$ an average over various row index $j$ stays approximately constant at $2$ within the time of internal wave propagation. As indicated by the time-space plot shown in Fig.\,\ref{fig:rx}(b), the fact that $r_{x}$ stays predominately $>1$ clearly illustrates the deficiency of momentum transfer due to wetting. Qualitatively, the above analysis demonstrates that the suppression of standing waves in vibrofluidized wet granular materials can be attributed to the strong tendency of collective motion along the driving direction and consequently the deficiency in momentum transfer. 

\begin{figure}
\centering
\resizebox{0.85\hsize}{!}{\includegraphics{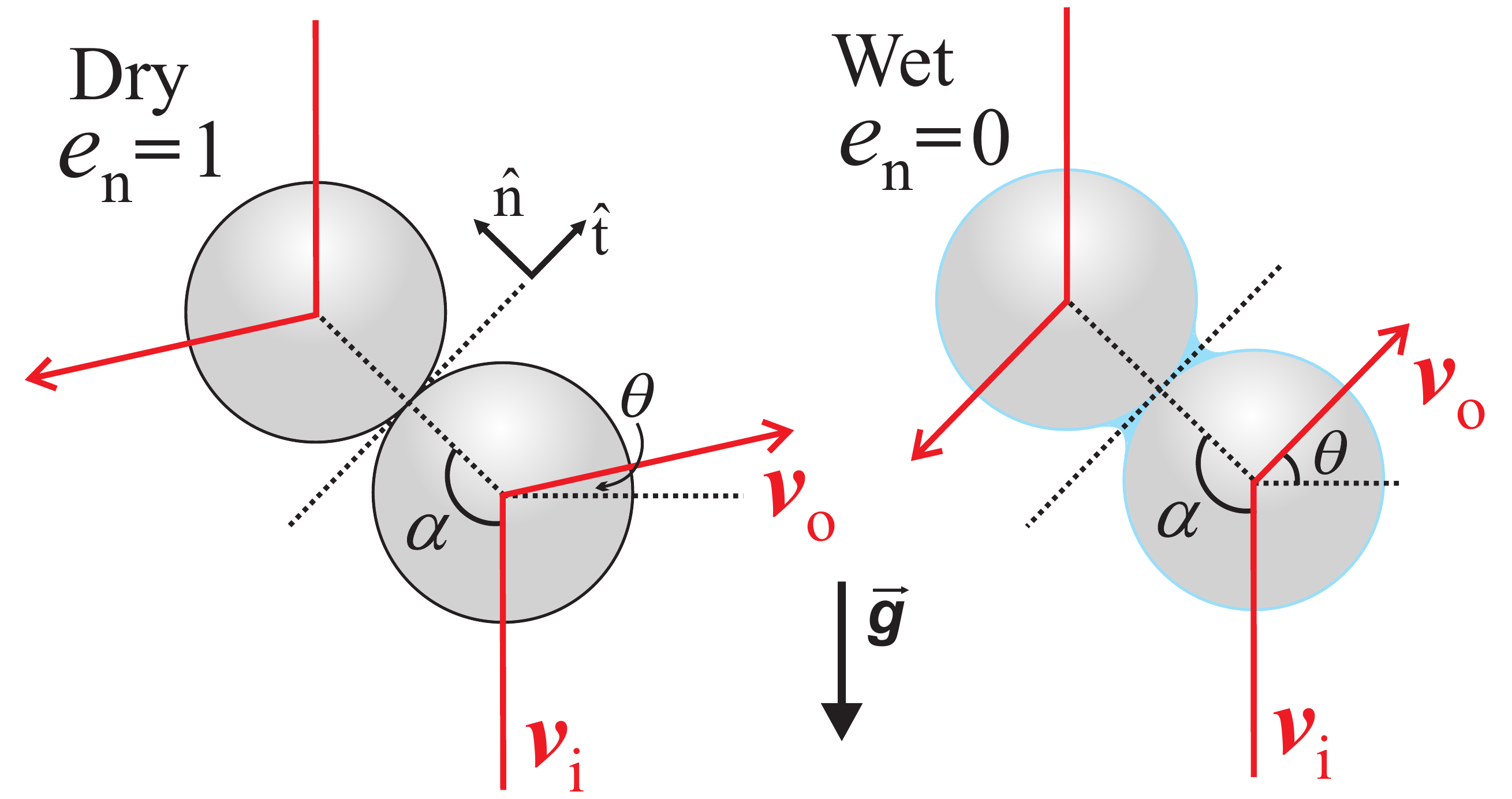}}
\caption{(Color online) A sketch to compare the momentum transfer from vertical to horizontal directions between two extreme (elastic and completely inelastic) cases of binary collisions with definitions of collision parameters.} 
\label{fig:model}       
\end{figure}

Quantitatively, the mean deficiency factor $\tilde{r}_{x} \approx 2$ can be understood from a comparison of momentum transfer in a binary oblique collision. For the sake of simplicity, I ignore the rotational degrees of freedom and consider the extreme case of $e_{\rm n} = 1$ and $0$ for the dry and wet cases (corresponding to the situation of $v_{\rm n} \to 0$), respectively. As sketched in Fig.\,\ref{fig:model}, two particles approach each other initially in the vertical direction with $\pi/2 < \alpha < \pi$. For the dry case, the rebound velocity $\vec v_{\rm o}$ is symmetric to $\vec v_{\rm i}$ along the normal direction $\hat{\vec n}$. Thus, the $x$ component of the rebound velocity for the lower particle yields $v_x^{\rm dry} = v_{\rm o} \cos \theta = v_{\rm i} \cos \theta = -v_{\rm i} \sin(2\alpha)$ with $\theta = 3\pi/2 - 2\alpha$. For the wet case, we have $\vec v_{\rm o}$ along the $\hat{\vec t}$ direction. Consequently, the corresponding $x$ component of the rebound velocity for the lower particle reads $v_x^{\rm wet} = v_{\rm o} \cos \theta = -v_{\rm i}\sin(2\alpha)/2 $, where $v_{\rm o} = v_{\rm i} \sin \alpha$ and $\theta = \pi - \alpha$. Therefore, the momentum transfer from the vertical to the horizontal direction for dry particles doubles that for wet ones in this ideal situation, independent of the colliding angle $\alpha$. Since the mean velocity of the two particles stays at $0$ for symmetry reasons, $\tilde{r}_{x} = v_x^{\rm dry}/v_x^{\rm wet}=2$, which agrees with the numerical result $2.17 \pm 0.56$ [see Fig.\,\ref{fig:rx}(a)] within the uncertainty. This argument, although idealized, quantitatively captures the deficiency in momentum transfer due to wetting.  

\section{Conclusion}
\label{sec:sum}

To summarize, this investigation demonstrates that the collective behavior of vibrofluidized granular materials can be tuned by the velocity dependent COR. Using an event driven algorithm, I show that standing waves in a thin layer of vibrofluidized granular material can be effectively suppressed via a change from dry to wet COR. Internal wave propagation, which accounts for the redistribution of the injected kinetic energy, is found to be weakly influenced by such a tuning. The suppression of standing waves arises from the different efficiency in momentum transfer from the vertical to the horizontal directions between dry and wet granular layers. Such a difference is further quantified with an momentum deficiency factor that compares the modulation of the velocity field in the horizontal direction between dry and wet particles. It is shown that the deficiency of momentum transfer due to wetting stays at about $2$, which can be rationalized with the momentum transfer at the level of individual particles. 

In the future, quantitative comparisons to experimental investigations are necessary to further develop and validate the model. Reciprocally, this approach is helpful in exploring the pattern creation mechanisms in wet granular materials from an insider view~\cite{Huang2011,Butzhammer2015,Zippelius2017}. As granular materials always have a certain size distribution, understanding how the polydispersity of particles influence the wave propagation also deserves further investigations. Moreover, it would also be interesting to explore how the redistribution of energy and momentum influences the convection induced granular capillary effect~\cite{Fan2017}. Last but not least, this investigation also paves the way of developing hybrid models that combine the advantages of ED and DEM simulations for modeling wet granular dynamics at a large scale.   

\begin{acknowledgments}
Helpful discussions with Ingo Rehberg and Simeon V\"olkel are gratefully acknowledged. This work is partly supported by the Deutsche Forschungsgemeinschaft through Grant No.~HU1939/4-1. 
\end{acknowledgments}

% BibTeX users please use
%\bibliographystyle{unsrt}
%\bibliography{habil_ext1}

\end{document}